\documentclass{article}
\usepackage{PRIMEarxiv}
\pdfoutput=1
\usepackage[T1]{fontenc}
\usepackage[utf8]{inputenc}
\usepackage{lmodern}
\usepackage{microtype}
\usepackage{url}
\usepackage[colorlinks=true,linkcolor=black,citecolor=black,urlcolor=black]{hyperref}

\usepackage[table]{xcolor}
\usepackage{amsmath,amsfonts,amssymb,mathrsfs}
\DeclareMathOperator*{\argmin}{arg\,min}
\DeclareFontFamily{OT1}{pzc}{}
\DeclareFontShape{OT1}{pzc}{m}{it}{<-> s * [1.10] pzcmi7t}{}
\DeclareMathAlphabet{\mathpzc}{OT1}{pzc}{m}{it}

\usepackage{graphicx}
\usepackage{epstopdf}
\DeclareGraphicsExtensions{.pdf,.eps,.png,.jpg,.mps}
\usepackage{booktabs}
\usepackage{multirow}
\usepackage{adjustbox}
\usepackage{stfloats}
\usepackage{float}

\usepackage[tight,footnotesize]{subfigure}

\usepackage{algorithm}
\usepackage{algorithmic}

\usepackage{bbm}
\usepackage{bm}
\usepackage{enumerate}
\usepackage{enumitem}
\usepackage[compress]{cite}
\usepackage{balance}
\usepackage{verbatim}

\usepackage{siunitx}
\makeatletter
\def\user@resume{resume}
\def\user@intermezzo{intermezzo}
\newcounter{previousequation}
\newcounter{lastsubequation}
\newcounter{savedparentequation}
\makeatother

\title{Range Asymmetric Numeral Systems-Based Lightweight Intermediate Feature Compression for Split Computing of Deep Neural Networks}

\author{
Mingyu~Sung, Suhwan~Im, Vikas Palakonda and Jae-Mo Kang\\
\textit{Department of Artificial Intelligence, Kyungpook National University, Daegu 41566, South Korea}\\
\texttt{alsrb0351@knu.ac.kr; tnghks9317@knu.ac.kr; vikas11475@knu.ac.kr; jmkang@knu.ac.kr}
}

\begin{document}
\maketitle

\begin{abstract}
Split computing distributes deep neural network inference between resource-constrained edge devices and cloud servers but faces significant communication bottlenecks when transmitting intermediate features. To this end, in this paper, we propose a novel lightweight compression framework that leverages Range Asymmetric Numeral Systems (rANS) encoding with asymmetric integer quantization and sparse tensor representation to reduce transmission overhead dramatically. Specifically, our approach combines asymmetric integer quantization with a sparse representation technique, eliminating the need for complex probability modeling or network modifications. The key contributions include: (1) a distribution-agnostic compression pipeline that exploits inherent tensor sparsity to achieve bandwidth reduction with minimal computational overhead; (2) an approximate theoretical model that optimizes tensor reshaping dimensions to maximize compression efficiency; and (3) a GPU-accelerated implementation with sub-millisecond encoding/decoding latency. Extensive evaluations across diverse neural architectures (ResNet, VGG16, MobileNetV2, SwinT, DenseNet121, EfficientNetB0) demonstrate that the proposed framework consistently maintains near-baseline accuracy across CIFAR100 and ImageNet benchmarks. Moreover, we validated the framework’s effectiveness on advanced natural language processing tasks by employing Llama2 7B and 13B on standard benchmarks such as MMLU, HellaSwag, ARC, PIQA, Winogrande, BoolQ, and OpenBookQA, demonstrating its broad applicability beyond computer vision. Furthermore, this method addresses a fundamental bottleneck in deploying sophisticated artificial intelligence systems in bandwidth-constrained environments without compromising model performance.
\end{abstract}


\section{Introduction}\label{sec:intro}
Deep neural networks (DNNs) have revolutionized artificial intelligence across diverse domains, achieving unprecedented performance in computer vision \cite{dosovitskiy2021image}, natural language processing \cite{lewis2020bart}, and multimodal learning \cite{wang2022image}. By stacking numerous layers—potentially in the hundreds—modern DNNs can capture intricate dependencies in high-dimensional data, achieving state-of-the-art (SOTA) accuracy on tasks as diverse as image recognition, language translation, and speech synthesis \cite{borisov2022deep}. However, this increased capability comes at substantial computational and memory costs, creating significant barriers to deployment on resource-constrained mobile devices.

These issues become particularly acute in \emph{edge computing} contexts, where local devices---such as smartphones, embedded sensors, and wearable devices---must operate under strict constraints~\cite{chang2021survey}. 
Two conventional paradigms exist for DNN inference in such settings:
\begin{enumerate}[leftmargin=1.5em]
    \item \textbf{Fully on-device inference}: 
    All computations (forward pass, intermediate processing, etc.) occur locally on the edge device. While this approach avoids data transfer overhead and can address privacy concerns, it often demands more memory and computational resources than many edge devices can provide~\cite{zheng2025review, walia2023ai}.
    Several recent studies explore fully on-device methods to reduce communication overhead and improve data privacy.
    Examples include model compression and quantization schemes, yet these can still be limited by the hardware capabilities of resource-constrained devices, leading to suboptimal performance in practice~\cite{zhao2024atom,sandler2018mobilenetv2,kim2022basq,shao2023omniquant}.

    \item \textbf{Fully off-device (cloud-based) inference}:
    The device offloads input data and all subsequent processing to a remote server in the cloud, minimizing local computation. However, this strategy does not fully utilize the growing computing potential of modern mobile devices. Furthermore, as the number of concurrent devices increases, it scales poorly—potentially overwhelming the server with excessive requests. In addition, high communication latency and bandwidth constraints become problematic in time-critical scenarios, and data security or privacy issues may arise when raw data is transmitted over unreliable networks.
\end{enumerate}
\noindent
Many edge platforms, therefore, struggle to maintain low-latency DNN execution under these two paradigms. 
They generally have limited memory capacities and relatively slow processors, making it prohibitively expensive (both computationally and energetically) to run large DNNs entirely on-device. At the same time, applications such as augmented reality and autonomous navigation impose strict latency requirements (on the order of milliseconds), making fully cloud-based solutions infeasible for real-time performance. As a result, scaling up SOTA DNNs on edge hardware remains a significant challenge.

\subsection{Related Works and Motivations}\label{sec:related_works_and_motivations}

\noindent
\textbf{Split computing.} 
Split computing (SC)~\cite{kang2017neurosurgeon} has emerged as a promising middle ground for balancing on-device computational limits and communication overhead. In this paradigm, only the initial portion of a DNN, often responsible for feature extraction, is deployed locally on the edge device, while the deeper layers are offloaded to the cloud. This division offers two key benefits: (i) reduced on-device computation compared to running the entire model locally, and (ii) lower data-transfer costs relative to sending raw inputs (e.g., high-resolution images or audio streams) over the network~\cite{yun2022cooperative,sung2025deco}. Most existing SC research~\cite{yun2022cooperative, sung2025deco, bakhtiarnia2023dynamic, cohen2020lightweight, oh2023communication} has focused on \emph{image-based} tasks, demonstrating compelling results in scenarios where partial inference can be performed efficiently on the edge before transmitting intermediate features (IFs) to the cloud.

\textbf{Existing approaches and limitations.}
Although previous works have addressed the compression of IFs in SC using various lightweight autoencoders or specialized network bottlenecks~\cite{duan2022efficient,matsubara2022supervised,muckley2023improving}, 
they often rely on explicit distribution modeling or task-specific architectures that may be difficult to adapt across different modalities (e.g., images, text) and diverse data distributions. Moreover, these methods typically focus on the feature representations rather than the encoding-decoding mechanism. Hence, while IF-level compression can be effective, the lack of a general, modality-agnostic approach to encoding--decoding in SC can hinder broader applicability---especially as data types and model architectures rapidly evolve.

\textbf{Gaps in current SC research.}
Despite promising progress on compressing feature tensors (such as learned embeddings), relatively few studies have thoroughly examined the \emph{entropy-coding perspective} of how these features are serialized, transmitted, and reconstructed. The use of sophisticated entropy coders like Range Asymmetric Numeral Systems (rANS)~\cite{duda2013asymmetric} has shown potential for high-fidelity compression but typically depends on carefully maintained symbol-frequency statistics, which may be computationally expensive or inflexible when adapting to new data domains (e.g., text vs.\ images). Additionally, most SC research has focused on vision tasks, whereas other modalities like text remain less explored~\cite{bajpai2023splitee, ohta2023lambda}. 
Given the proliferation of large language models (LLMs) in domains such as text generation, summarization, and dialogue systems, transmitting and processing IFs of textual data is fast becoming as critical---if not more so---than handling large images~\cite{cheng2024advancements}.

\noindent
\textbf{Motivation.}
Recognizing these gaps, our work emphasizes the encoding--decoding pipeline itself rather than relying solely on specialized feature-extraction techniques. 
Specifically, we seek a compression method that:
\begin{itemize}[leftmargin=1.2em]
    \item \textit{Minimizes the reliance on distribution modeling}, allowing simpler and more robust deployments in dynamic or unknown data regimes,
    \item \textit{Achieves consistently high compression ratios and low latency} across different tasks---extending beyond standard computer-vision scenarios to encompass LLM-based language tasks,
    \item \textit{Maintains general applicability}, integrating seamlessly with various DNN architectures (CNNs, Transformers, etc.) and handling diverse modalities (images, text) without substantial overhead.
\end{itemize}
These requirements underscore the need for a direct and lightweight encoding--decoding approach that does not hinge on complex probability estimation or domain-specific bottlenecks, thereby paving the way for scalable SC solutions in computer vision and natural language processing. 

\subsection{Contributions}\label{sec:contributions}

\noindent
\textbf{Overview of our approach.}
In this paper, we propose a novel compression pipeline based on rANS that operates directly on quantized and sparse representations of IFs, irrespective of the input domain or underlying neural architecture. Our method avoids explicit probability modeling, instead leveraging integer quantization and a lightweight sparse-format encoding to achieve efficient compression. This design proves advantageous for various applications and data modalities, aligning well with the emerging need to support image-based and language-oriented tasks under limited-resource settings.

The main contributions of the paper are highlighted as follows:

\begin{itemize}[leftmargin=1.6em]
    \item \textbf{Direct rANS compression without distribution modeling.} We eliminate the overhead of building or updating explicit probability tables by reformatting the IF tensor into a sparse, integer-only representation. This makes our pipeline robust to evolving data distributions, including those encountered in text-based applications with LLMs.
    
    \item \textbf{GPU-optimized pipeline for real-time encoding and decoding.} Through careful algorithmic design and data-structure choices (e.g., modified compressed sparse row encoding), our framework harnesses GPU parallelism to achieve low-latency compression. This ensures its suitability for latency-critical edge--cloud systems, regardless of whether the data is image or text based.
    
    \item \textbf{Broad applicability across tasks and models.} The proposed method is \emph{architecture-agnostic} and readily integrates into a wide range of DNN topologies. As a result, it seamlessly extends to both classical convolutional networks and the rapidly evolving family of transformer-based language models, maintaining strong compression ratios and minimal accuracy loss.
\end{itemize}

The remainder of this paper is organized as follows. 
Section~\ref{sec:prelim} provides essential background on split computing and reviews the fundamentals of rANS encoding. 
Section~\ref{sec:proposed} details our proposed compression pipeline, highlighting the roles of sparsity and integer quantization in achieving an efficient implementation. 
In Section~\ref{sec:experiments}, we present comprehensive experimental results on both image-centric benchmarks (e.g., CIFAR100, ImageNet) and preliminary tests with LLMs, demonstrating consistent gains in compression and runtime efficiency. 
Finally, Section~\ref{sec:conclusion} concludes the paper and discusses future research avenues for extending our approach to broader scenarios.

\section{Preliminaries}\label{sec:prelim}
\begin{figure*}[htb!]
    \centering
    \includegraphics[width=\linewidth]{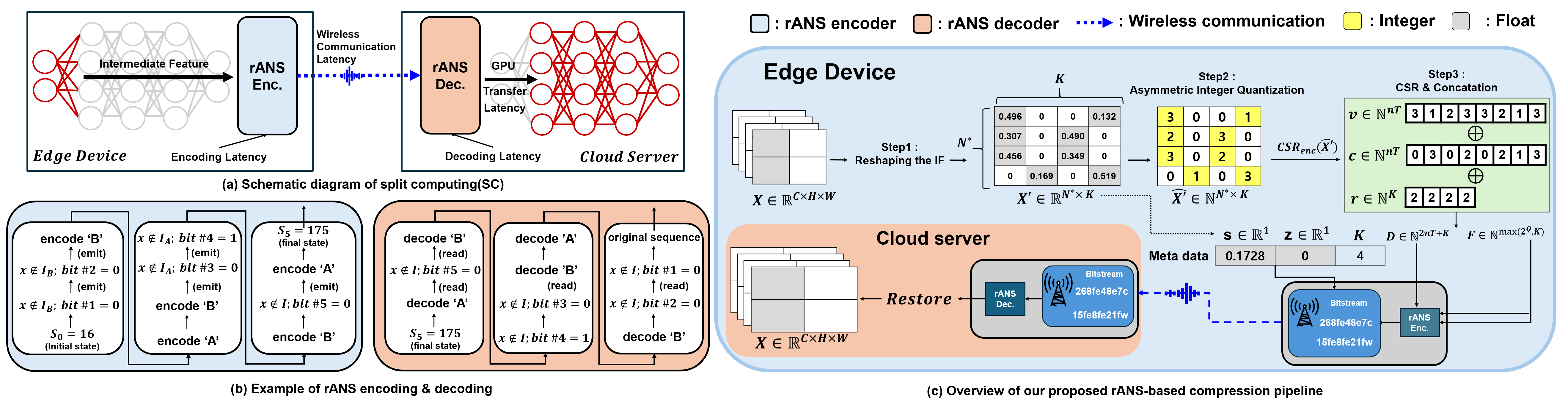}
    \caption{\textbf{(a)} Schematic diagram of split computing (SC). Due to limited memory on the edge device, only the initial layers of a DNN run locally, while the cloud server processes subsequent layers. IFs from the edge are compressed, transmitted over a wireless link, and decoded on the cloud side before final-layer inference. The four main latency contributors in SC are: (i) edge encoding, (ii) wireless transfer, (iii) cloud decoding, and (iv) GPU integration. \textbf{(b)} Illustrative example of rANS encoding \& decoding. Symbols (`A', `B') are successively encoded into a single state using rANS; decoding recovers the symbols by reversing these state transitions. Notation such as $s_i$ denotes the internal state after processing symbol $i$. \textbf{(c)} Overview of our proposed rANS-based compression pipeline. An IF tensor \(X \in \mathbb{R}^{C \times H \times W}\) is reshaped and quantized to produce integer symbols, which are then packed into a modified CSR format and concatenated into a single vector \(\mathbf{D}\). Finally, rANS encodes \(\mathbf{D}\) into a compact bitstream that is transmitted to the cloud and decoded prior to completing the final DNN layers.}
    \label{system_model}
\end{figure*}

\subsection{Brief Introduction to rANS}\label{sec:SC_rANS}
ANS is a family of entropy-coding methods designed to achieve near-optimal compression rates. Two main variants exist: table-based ANS (tANS) and rANS~\cite{duda2013asymmetric}. 
In tANS, encoding and decoding are driven by pre-computed lookup tables constructed from the symbol probability distribution~\cite{weissenberger2019massively}. This approach can be high-speed on GPUs~\cite{knorr2021ndzip,zhang2023fz}, but its lookup tables proliferate with the number of symbols and states. This can be problematic when the IFs must be transmitted over limited communication links, as large tables would add significant overhead. In contrast, rANS uses variable-length coding, meaning the bit-length assigned to each symbol depends on its relative frequency. Because rANS compresses symbols according to their probabilities, it can approach the theoretical entropy limit.

\paragraph{Entropy Bound and Compression Ratio.}
Let $m$ be the number of unique symbols, each occurring $f(x_i)$ times (for $i = 1,2,\dots,m$) out of a total of $N$ symbols. The probability of symbol $x_i$ is $p(x_i) \;=\; \frac{f(x_i)}{N}$.
Then the expected compressed size $\eta$ and compression ratio $\rho$ can be written as
\begin{equation}
\eta \;=\; N \cdot H \;=\; -N \sum_{i=1}^{m} p(x_i)\log_2 p(x_i),
\quad
\rho \;=\; \frac{\eta}{N \,\log_2 \mathcal{A}},
\label{entropy}
\end{equation}
where $\mathcal{A}$ is the number of possible symbols in the data and $H$ is the Shannon entropy. Specifically, $\rho$ measures how closely the compressed representation approaches the ideal coding length $\log_2 \mathcal{A}$ per symbol and is a practical metric for compression efficiency.

\paragraph{rANS Coding Process.}
Given a symbol $x_i$ with frequency $f(x_i)$ and its CDF value $F(x_i)$, rANS transforms a previous state $s_{i-1}$ into a new state $s_i$:
\begin{equation}
s_i \;=\; \left\lfloor \frac{s_{i-1}}{f(x_i)} \right\rfloor 
          \cdot 2^n 
          \;+\; F(x_i) 
          \;+\; \bigl(s_{i-1} \bmod f(x_i)\bigr),
\label{rans_encoding}
\end{equation}
where $n$ is the number of bits associated with the coding precision (i.e., $2^n$ scales the state space).
Decoding is the inverse process: from a current state $s_i$, we identify the symbol $x_i$ such that
\begin{equation}
F(t) \;\le\; \bigl(s_i \bmod 2^n\bigr) < F(t+1),
\label{rans_symbol_selection}
\end{equation}
and recover $x_i$. The value $s_i$ is then updated accordingly by reversing~\eqref{rans_encoding}:
\begin{equation}
s_{i-1} \;=\; f(x_i)\,\left\lfloor \frac{s_i}{2^n}\right\rfloor 
              \;+\; \bigl(s_i \bmod 2^n\bigr) \;-\; F(x_i).
\end{equation}

\paragraph{Renormalization and Practical Details.}
A key aspect of rANS is renormalization, which ensures the internal state remains within a safe range to preserve coding precision without overflow or underflow. Concretely, if $s_{i-1}$ is too small or too large for the upcoming symbol, a portion of its bits is emitted (in encoding) or fetched from the bitstream (in decoding) such that $s_{i-1}$ is adjusted back into a predefined interval (e.g., $[2^n,\,2^{n+1}-1]$). This guarantees that the integer divisions and moduli in~\eqref{rans_encoding} behave consistently:
\begin{itemize}
    \item \textbf{Encoder Side:} If $s_{i-1} < \mathrm{threshold}$, some bits from $s_{i-1}$ are flushed (written) to the output bitstream. After shifting $s_{i-1}$ left and emitting the lower bits, the resulting enlarged state is used for~\eqref{rans_encoding}.
    \item \textbf{Decoder Side:} Conversely, whenever $s_i$ is below the threshold, bits are read from the compressed bitstream to shift $s_i$ into the valid range before applying the inverse of~\eqref{rans_encoding}.
\end{itemize}
In practice, this procedure is symbol-dependent if frequencies $f(x_i)$ are uneven. For example, if a particular symbol $x_i$ has a tiny $f(x_i)$, one might require a larger minimal $s_{i-1}$ to avoid excessive writes or reads. This can be implemented using conditionals such as ``If $s_{i-1} < \mathrm{limit}[x_i]$, emit/read bits.'' The sets $I_x$ often seen in rANS code examples refer to valid ranges of the state $s$ in which symbol $x$ can be encoded without extra normalization steps.

To illustrate these mechanics in detail, Fig.~\ref{system_model}(b) presents a step-by-step example with two symbols, `A' and `B'.  
The figure highlights how $s_{i-1}$ is updated (via divisions, moduli, and bit-shifts) at each encoding step and how renormalization flushes or retrieves bits to keep the internal state in range. 
Likewise, the decoding process reverses these transformations, demonstrating how rANS achieves efficient entropy coding by carefully managing the evolution of $s_i$.

\paragraph{Key Observations}
\begin{itemize}[leftmargin=1.2em]
\item \emph{Small Alphabet or Highly Skewed Frequencies:}
      If the data has only a few unique symbols or very imbalanced symbol frequencies, the entropy $H$ is typically low, allowing for high compression ratios.
\item \emph{Uniform Distributions:}
      If symbols are equally likely, the entropy is high, and potential compression gains are limited.
\item \emph{Rare Symbols:}
     Symbols with very low $f(x_i)$ can cause frequent renormalizations, and bitstream writes, increasing overhead and slowing down the encoder/decoder. Appropriate choice of $n$ (precision) or a careful grouping of rare symbols may alleviate this.
\end{itemize}

\noindent
Overall, rANS balances efficient entropy coding and practical implementation on modern hardware, making it a strong candidate for compressing IFs (e.e., feature maps in a neural network pipeline) in SC scenarios. By carefully managing state ranges and renormalizing only when needed, rANS achieves near-optimal compression rates with minimal memory overhead compared to table-based approaches. Future work includes optimizing renormalization thresholds to reduce bitstream I/O further and investigating specialized GPU kernels for highly parallel rANS~\cite{knorr2021ndzip,zhang2023fz}.

\subsection{System model}\label{sec:SC_concept}
\noindent
In many edge--cloud scenarios, memory constraints on the edge device prevent an entire DNN from running locally. As depicted in Fig.~\ref{system_model}(a), only the initial layers of the DNN (e.g., feature extraction) are executed on the resource-limited edge. In contrast, the subsequent layers (e.g., higher-level inference tasks) run on a cloud server equipped with ample memory and processing capacity.
Once the initial layers finish on the edge, the resulting IFs are compressed and transmitted over a wireless link\footnote{Real-world SC deployments may encounter unreliable transmission channels or packet loss. Recent works~\cite{sung2025deco,yun2022cooperative} have therefore introduced concepts like \emph{e-outage} to optimize the chosen split layer under such conditions. Similarly, our method considers scenarios with e-outage to handle transmission uncertainties and mitigate packet loss.}. Upon arrival in the cloud, the compressed IFs are passed to a GPU and decoded to process the remaining DNN layers and complete the inference pipeline.

\vspace{6pt}
\noindent
\textbf{Key Latency Factors in SC.}
Four primary sources of latency arise in this pipeline:
\begin{itemize}[leftmargin=1.5em]
    \item \emph{Edge-side encoding} of the IFs, which can introduce a computational overhead if the compression algorithm is complex;
    \item \emph{Wireless transmission} of the compressed bitstream, where the bitstream size directly affects transfer time;
    \item \emph{Cloud-side decoding}, which must be efficient to avoid bottlenecks upon arrival;
    \item \emph{GPU data transfer and final-layer computation}, including the the overhead of loading the reconstructed IFs into the GPU’s memory.
\end{itemize}
A smaller bitstream reduces the transmission delay but may demand more complex encoding or decoding steps; conversely, a larger bitstream can overload the link and slow the data transfer into GPU memory. 
Therefore, an effective compression strategy must balance these competing factors to keep total latency low. 
Our proposed method seeks precisely this balance by minimizing the bitstream size (cutting transmission time) and ensuring that both the encoder and decoder remain computationally efficient (avoiding excessive delays on either side).

\section{Proposed Method}\label{sec:proposed}

\subsection{rANS-based Compression Pipeline}\label{sec:pipeline}
Fig.~\ref{system_model} (c) depicts an overview of our proposed pipeline, which comprises four main steps: (i) reshaping the IF tensor to control its entropy and improve compressibility, (ii) quantizing the reshaped data using asymmetric integer quantization (AIQ) for reduced precision and computational cost, (iii) encoding nonzero entries via a modified Compressed Sparse Row (CSR) format to exploit sparsity, and (iv) unifying all encoded symbols in a single concatenated vector for rANS compression on the GPU with minimal data transfers.
Building on the SC paradigm, these steps reduce communication overhead while maintaining near-baseline accuracy.

\paragraph{Reshaping the IF}
Let the IF tensor be \(X \in \mathbb{R}^{C \times H \times W}\). 
We reshape \(X\) into \(X' \in \mathbb{R}^{N \times K}\) according to
\begin{equation}
\label{eq1}
X \in \mathbb{R}^{C \times H \times W} 
\;\longrightarrow\; 
X' \in \mathbb{R}^{N \times K},
\quad
N = \tfrac{T}{K}, 
\quad
T = C \times H \times W.
\end{equation}
Choosing appropriate \((N,K)\) can substantially impact the statistical distribution of compressed IFs and thus the overall compression efficiency. 
The reshaping step is a key \emph{optimization objective} of our method, as it ultimately modifies the entropy profile of the data.

\paragraph{Asymmetric Integer Quantization (AIQ)}
Before applying CSR encoding, we quantize \(X'\) into integer form:
\begin{equation}
\label{eq:AIQ_main}
\hat{x} 
\;=\; 
\left\lfloor \dfrac{x}{s} + z \right\rceil,
\quad
s \;=\; \dfrac{x_{\max}-x_{\min}}{2^Q - 1},
\quad
z \;=\;\left\lfloor -\dfrac{x_{\min}}{s} \right\rceil,
\end{equation}
where \(x\) is an element of \(X'\), \(Q\) is the bit-width (e.g., 4 or 8), and \(\lfloor \cdot \rceil\) indicates rounding to the nearest integer. Consequently, each \(\hat{x}\) lies in \(\{0, \dots, 2^Q - 1\}\). Because this integer-only representation avoids floating-point operations, it achieves higher compression ratios and accelerates subsequent GPU processing on constrained devices.

\paragraph{Modified CSR Encoding}
Many DNNs produce IFs containing a substantial fraction of zeros. We define \(n\) as the number of nonzero elements in \(X'\). Storing only these \(n\) nonzero values can significantly reduce data transfer overhead. To capitalize on this sparsity, we encode \(\hat{X}\) (the quantized IF) in a \emph{modified} CSR format. Unlike the standard CSR, where \(\mathbf{r}[i]\) is the cumulative count of nonzeros up to row \(i\), we use a \emph{non-cumulative} scheme in which \(\mathbf{r}[i]\) is the direct count of nonzeros in row \(i\). By deferring cumulative-sum computation until \emph{decoding}, we diminish the dynamic range of symbols, leading to more efficient rANS coding. Concretely, we construct three arrays:
\begin{itemize}[leftmargin=1.2em]
    \item \(\mathbf{v} \in \mathbb{N}^{nT}\): nonzero values (post-quantization),
    \item \(\mathbf{c} \in \mathbb{N}^{nT}\): column indices of the nonzeros,
    \item \(\mathbf{r} \in \mathbb{N}^{N}\): row-wise counts of nonzeros (non-cumulative).
\end{itemize}
Converting from dense to modified CSR has time complexity \(\mathcal{O}(T)\), necessitating only a single pass over \(X'\). 
This design choice directly improves rANS efficiency by reducing symbol variance.

\paragraph{Concatenation and rANS Encoding}
Having obtained \(\mathbf{v},\mathbf{c},\mathbf{r}\), we concatenate them into a single vector
\[
\mathbf{D} = \mathbf{v} \;\oplus\; \mathbf{c} \;\oplus\; \mathbf{r},
\]
whose length is \(\ell_{\mathbf{D}} = 2nT + N\). 
This unification step minimizes memory transfers between CPU and GPU, since we can load, compress, and transfer \(\mathbf{D}\) as a single block rather than handling multiple structures. 
It also ensures that rANS can be efficiently applied in a single pass on the GPU, effectively reducing GPU transfer latency and simplifying synchronization.

To perform rANS encoding, we construct frequency vectors for \(\mathbf{v},\mathbf{c},\mathbf{r}\), pad them to a uniform length, and sum them elementwise to form \(\mathbf{F}\). 
Because \(\mathbf{v}\) spans \(\{0,\dots,2^Q-1\}\), while \(\mathbf{c}, \mathbf{r}\) may include indices up to \(K\), the overall alphabet for rANS is $\max(2^Q-1, K)$.

\subsection{Approximate Theoretical Cost Model for rANS}\label{sec:analysis}
\noindent
\textbf{An Optimal Reshape Dimension.}
The reshaping step does more than merely rearrange the data: it changes the frequency distribution of symbols in $\mathbf{D}$, affecting both the entropy and the achievable compression ratio. In particular, the distributions of \(\mathbf{c}\) (column indices) and \(\mathbf{r}\) (row counts) exert a strong influence on entropy. Concretely, after reshaping \(\mathbf{X}\in\mathbb{R}^{C\times H \times W}\) to \(\mathbf{X}' \in \mathbb{R}^{N \times K}\) (with \(K = T/N\) and \(T=C\!\times H \times W\)), the CSR-based encoding yields three arrays \(\mathbf{v}, \mathbf{c}, \mathbf{r}\). Their combined frequency distribution \(\mathbf{F}\) determines the Shannon entropy \(H\bigl(\mathbf{p}(N)\bigr)\). A well-chosen \(N\) skews \(\mathbf{F}\), i.e., it increases the probability mass on a smaller subset of symbols, thereby reducing entropy. By contrast, a poor choice in reshaping can flatten \(\mathbf{F}\) and yield a higher entropy. Consequently, finding an optimal \(N\) is paramount for minimizing rANS’s total cost (including encoding, decoding, and communication).

\begin{figure}[htbp]
    \centering
    \includegraphics[width=0.80\linewidth]{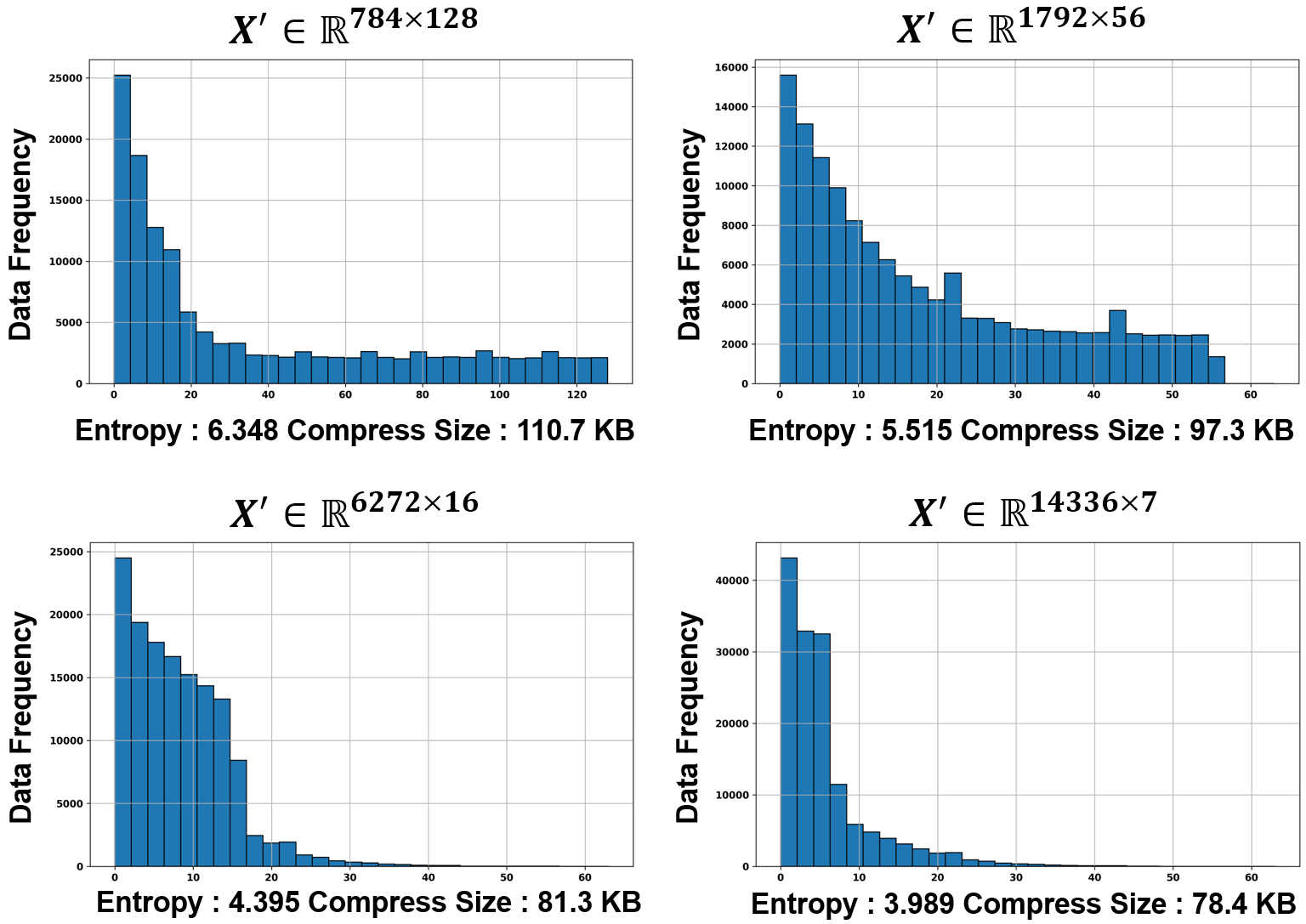}
    \caption{Illustration of how reshaping an IF \(X \in \mathbb{R}^{128\times 28\times 28}\) affects the data distribution and entropy, ultimately impacting the compressed size. Each subfigure corresponds to reshaping \(X\) into \(\mathbb{R}^{784\times 128}\), \(\mathbb{R}^{1792\times 56}\), \(\mathbb{R}^{6272\times 16}\), and \(\mathbb{R}^{14336\times 7}\), respectively. The histograms depict how the frequency distribution of unique values (post-quantization) shifts with different reshape dimensions, while the reported entropies and compressed sizes underscore the correlation between a more skewed distribution and improved compression.}
    \label{fig:rans_cost_model}
\end{figure}

\noindent
\textbf{Why Reshaping Matters.}
Fig.~\ref{fig:rans_cost_model} illustrates the impact of four different reshape configurations on an example IF \(X \in \mathbb{R}^{128 \times 28 \times 28}\). When \(X\) is reshaped to \(\mathbb{R}^{784 \times 128}\), the entropy is relatively high (6.348), leading to a larger compressed size (110.7\,KB). As \(N\) grows and \(K\) shrinks (e.g., \(\mathbb{R}^{14336 \times 7}\)), the distribution becomes more skewed, lowering the entropy to 3.989 and reducing the compressed size to 78.4\,KB. This empirical observation reinforces the idea that carefully choosing \((N,K)\) can effectively skew the symbol distribution, cutting the Shannon entropy and thus improving rANS compression.

\noindent
\textbf{From Skewness to rANS Efficiency.}
Methods based on arithmetic coding, including rANS, approach the theoretical Shannon limit. Consequently, a skewed distribution---one in which a small subset of symbols occurs with high probability---naturally lowers the average code length. Formally, let \(\mathbf{p}(N)\) denote the probability distribution of symbols in \(\mathbf{D}\). The expected bit-rate under rANS is Eq.~\eqref{entropy}. If reshaping drives more probability mass onto fewer symbols, it reduces the entropy and, thus, the expected bits per symbol.

\noindent
\textbf{Skew-Inducing Reshaping.}
A reshaping from \(X\) to \(X'\) modulates how nonzeros are distributed across rows and columns:
\begin{itemize}[leftmargin=1.5em]
    \item \emph{Row counts \(\mathbf{r}\).}
    Increasing \(N\) (thus reducing \(K\)) tends to distribute nonzeros across more rows but fewer columns, concentrating the row-count distribution \(p(r)\). Conversely, smaller \(N\) (larger \(K\)) can spread nonzeros more evenly, flattening \(p(r)\).
    \item \emph{Column indices \(\mathbf{c}\).}
    Because \(\mathbf{c}\) encodes column positions, its distribution \(p(c)\) depends on \(K\). A smaller \(K\) narrows the range \(\{0,\dots,K-1\}\), which can significantly reduce entropy if nonzeros concentrate in specific columns.
    \item \emph{Nonzero values \(\mathbf{v}\).}
    Although \(\mathbf{v}\) (the quantized magnitudes) is less sensitive to \(N\) than \(\mathbf{c}\) or \(\mathbf{r}\), certain reshapes can align channel- or spatial-wise correlations, further skewing \(p(v)\).
\end{itemize}

\subsection{Cost Model and Objective Function}
\begin{figure}[htb!]
\centering
\includegraphics[width=0.82\linewidth]{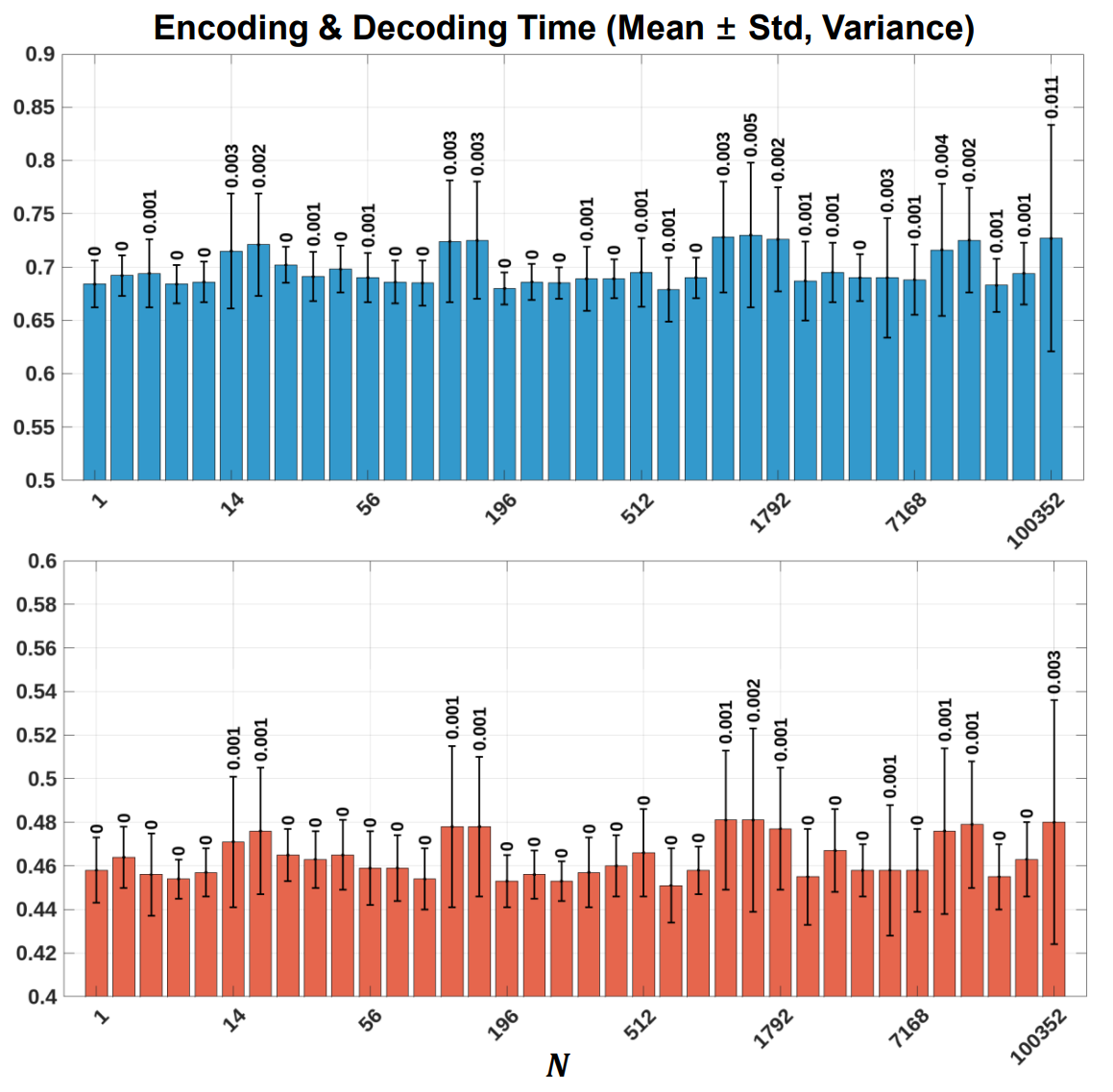}
\caption{Measured \(\alpha_{\mathrm{enc}} \cdot T_{\mathrm{enc}}(N)\) (top) and \(\alpha_{\mathrm{dec}} \cdot T_{\mathrm{dec}}(N)\) (bottom) in milliseconds, as a function of the reshape dimension \(N\). Despite varying \(N\) over several orders of magnitude, both operations exhibit nearly constant runtimes, indicating that GPU-based parallel processing keeps overhead low in practice. Error bars denote standard deviations across multiple trials.}
\label{fig:approx_vs_actual}
\end{figure}

To formalize this trade-off, we define the total cost of encoding, decoding, and communication via:
\begin{equation}
\label{eq:total_cost}
\begin{aligned}
& N^{\ast} = \argmin_{N \in \{1,\dots,T\}}\; T(N), \text{where} \\[4pt]
& \quad T(N) 
  \;=\;
  \alpha_{\mathrm{enc}}T_{\mathrm{enc}}(N) 
  \;+\;
  \alpha_{\mathrm{dec}}T_{\mathrm{dec}}(N) 
  \;+\;
  T_{\mathrm{tot}}(N), \\[5pt]
& \quad T_{\mathrm{tot}}(N) 
  \;=\;
  \ell_{\mathbf{D}}\,H\bigl(\mathbf{p}(N)\bigr).
\end{aligned}
\end{equation}
Here, \(\alpha_{\mathrm{enc}}\) and \(\alpha_{\mathrm{dec}}\) denote empirical constants indicating how encoding and decoding times scale with symbol count, and $\ell_{\mathbf{D}}$ denotes the total symbol count after CSR. The term \(H\bigl(\mathbf{p}(N)\bigr)\) is the Shannon entropy of the reshaped distribution.

\paragraph{Encoding and Decoding in Practice.}
Although \eqref{eq:total_cost} accounts for encoding and decoding latencies, in the range of data sizes we consider, these steps involve relatively lightweight operations that are highly parallelizable on GPUs, keeping \(T_{\mathrm{enc}}(N)\) and \(T_{\mathrm{dec}}(N)\) nearly invariant with respect to \(N\). 
As shown in Fig.~\ref{fig:approx_vs_actual}, even when \(N\) varies over several orders of magnitude, the measured runtimes for both encoding (top) and decoding (bottom) remain almost constant.
Consequently, these components have negligible influence on determining the optimal reshape dimension \(N\).
The dominant factor in~\eqref{eq:total_cost} is the entropy-dependent term \(T_{\mathrm{tot}}(N)\), which characterizes the amount of information that must be transmitted after compression.
Strictly speaking, \(T_{\mathrm{tot}}(N)\) is not the actual latency but rather a proxy proportional to the bitstream size.
Therefore, minimizing \(H\bigl(\mathbf{p}(N)\bigr)\) directly improves compression efficiency and overall transmission cost.

\paragraph{Constrained Approximate Search for a Near-Optimal \(\Tilde{N}\).}
While exhaustive enumeration over all \(N \in \{1,\dots,T\}\) would yield the true optimum \(N^{\ast}\), it is computationally expensive. 
We thus constrain the feasible search domain and employ an early-stopping heuristic:
\begin{enumerate}[leftmargin=1.1em, itemsep=2pt, topsep=3pt]
  \item \textbf{Enforce \(N > K\).}  
  Since \(K = T/N\), requiring \(N > K\) implies \(N^2 > T\). This leads us to skip any \(N \le \sqrt{T}\).  In many CSR-based applications, this restriction is justified by the need for more rows than columns to preserve row-compression benefits.

  \item \textbf{Restrict \(K \leq 2^Q\).}  
  Rewriting \(K \le 2^Q\) as \(\frac{T}{N} \le 2^Q\) yields \(N \ge \frac{T}{2^Q}\).  Thus, we skip values of \(N\) below \(\bigl\lceil T / 2^Q \bigr\rceil\).  Allowing \(K\) to exceed \(2^Q\) would significantly inflate the effective alphabet, undermining compression efficiency.
  \item \textbf{Early stopping when \(T_{\mathrm{tot}}(N)\) stops decreasing.}
  By iterating \(N\) in descending order and halting once \(T_{\mathrm{tot}}(N)\) grows relative to the prior iteration, we typically identify a near-optimal \(\Tilde{N}\) with only a fraction of the full search.
\end{enumerate} Algorithm~\ref{alg:gpu_edge_enumeration} consolidates the domain-pruning rules and early-stopping mechanism into a practical procedure for finding a near-optimal reshape dimension. Specifically, $N^*$ denotes the global optimum obtained by exhaustively testing all valid $N\in\{1,\dots,T\}$, whereas $\Tilde{N}$ is the best solution found under our constraints. Despite not searching the entire range of $N$, the early-stopping heuristic typically provides a solution $\Tilde{N}$ that closely approximates $N^*$ while significantly reducing search time.

\noindent

\begin{algorithm}[htb!]
\caption{Approximate Enumeration for Optimal \(\Tilde{N}\)}
\label{alg:gpu_edge_enumeration}
\begin{algorithmic}[1]
\REQUIRE 
  \(\mathbf{X} \in \mathbb{R}^{T}\), 
  \(\alpha_{\mathrm{enc}} = 0\), \(\alpha_{\mathrm{dec}} = 0\), 
  \(Q\), \(n\)
\ENSURE 
  \(\Tilde{N} = \arg\min_{N}T_{\mathrm{tot}}(N)\)
\STATE \textbf{Domain Restrictions:}
\[
   N > \sqrt{T}, \quad \frac{T}{N} \le 2^Q, \quad N \in \mathbb{N}.
\]
\STATE \textbf{Compute bounds:}
\[
   N_{\min} \gets \max\!\Bigl(\lfloor \sqrt{T} \rfloor + 1,\; \lceil T / 2^Q \rceil\Bigr),\quad
   N_{\max} \gets T.
\]
\STATE Initialize:
\[
   \begin{aligned}
      \texttt{best\_cost} &\gets +\infty,\quad \Tilde{N} \gets N_{\max},\\
      \texttt{prev\_cost} &\gets +\infty,\quad \texttt{decreasing} \gets \texttt{True}.
   \end{aligned}
\]
\FOR{\(N = N_{\max}\) \textbf{downto} \(N_{\min}\)}
  \IF{\(T \bmod N \neq 0\)} \STATE \textbf{continue} \COMMENT{Skip non-integer \(K=T/N\).} \ENDIF
  \STATE \(\mathbf{D},\,\mathbf{F} \gets \textsc{CSR\_Encode}\bigl(\mathbf{X},\,Q,\,\text{reshape} = [N,\,T/N]\bigr)\)
  \STATE \(H\bigl(\mathbf{p}(N)\bigr) \gets \textsc{ComputeEntropy}\bigl(\mathbf{F}\bigr)\)
  \STATE \(T_{\mathrm{tot}}(N) \gets \ell_{\mathbf{D}}\,H\bigl(\mathbf{p}(N)\bigr)\)
  \IF{\(T_{\mathrm{tot}}(N) < \texttt{best\_cost}\)} 
      \STATE \(\texttt{best\_cost} \gets T_{\mathrm{tot}}(N)\); \(\Tilde{N} \gets N\)
  \ENDIF
  \IF{\(\texttt{decreasing} \land (T_{\mathrm{tot}}(N) > \texttt{prev\_cost})\)} 
      \STATE \textbf{break} 
  \ENDIF
  \STATE \(\texttt{prev\_cost} \gets T_{\mathrm{tot}}(N)\)
\ENDFOR
\RETURN \(\Tilde{N}\)
\end{algorithmic}
\end{algorithm}

\paragraph{Complexity Analysis.}
Because we require both $N$ and $K=T/N$ to be integers, the algorithm only evaluates divisors of $T$. 
In the worst case, let $\mathcal{D}$ be the set of all divisors; then $|\mathcal{D}|$ can be as large as $\exp(O(\log T/\log\log T))$ in theory, but in practice it is much smaller than $T$. 
At each valid $N \in \mathcal{D}$, the procedure performs $O(T)$ work (primarily CSR encoding and entropy calculation), leading to a worst-case complexity of 
\[
O\bigl(|\mathcal{D}| \cdot T\bigr).
\]
Nonetheless, two factors significantly reduce average running time:
\begin{enumerate}[leftmargin=1.5em,itemsep=0pt,topsep=0pt]
    \item \emph{Domain Restrictions:} We skip any $N \le \sqrt{T}$ or violating $K \leq 2^Q$, drastically pruning the search space.
    \item \emph{Early Stopping:} Once $T_{\mathrm{tot}}(N)$ stops decreasing for consecutive iterations, the loop terminates. Because $T_{\mathrm{tot}}(N)$ typically decreases until it reaches a region close to its minimum and then rises, the algorithm converges quickly.
\end{enumerate}
Hence, although the formal bound remains $\mathcal{O}(|\mathcal{D}|\cdot T)$, the practical run time is often much smaller. As such, it strikes an attractive balance between computational cost and compression efficacy.

\section{Results and Discussion}\label{sec:experiments}
\subsection{Experimental Setup}
All experiments were conducted on a server with an AMD EPYC 7352 24-core CPU, 256\,GB of RAM, and an NVIDIA A6000 GPU (48\,GB VRAM), running Ubuntu 20.04 and CUDA 11.4. We evaluated our rANS-based framework on (i) vision tasks (CIFAR100 \cite{krizhevsky2009learning}, ImageNet \cite{deng2009imagenet}) using ResNet (34/50) \cite{he2016deep}, VGG16 \cite{simonyan2014very}, MobileNetV2 \cite{sandler2018mobilenetv2}, SwinT \cite{liu2021swin}, DenseNet121 \cite{huang2017densely}, and EfficientNetB0 \cite{tan2019efficientnet}, and (ii) NLP tasks (MMLU \cite{hendrycks2021mmlu}, HellaSwag \cite{zellers2019hellaswag}, ARC \cite{clark2018arc}, PIQA \cite{bisk2020piqa}, Winogrande \cite{sakaguchi2020winogrande}, BoolQ \cite{clark2019boolq}, OpenBookQA \cite{mihaylov2018openbookqa}) using Llama2 7B and 13B. We used off-the-shelf pre-trained weights without extra fine-tuning, and for each quantization level $Q\in\{2,3,4,6,8\}$, we measured data size, encoding/decoding times, and top-1 accuracy or task-specific metrics. For communication measurements, we adopted the \(\varepsilon\)-outage model from \cite{yun2022cooperative} to account for noisy wireless transmission latency, letting \(\displaystyle T_{\mathrm{comm}}(N)\) denote the resulting $\epsilon$-outage-based communication latency, and set \(\varepsilon=0.001\), \(W=10\,\mathrm{MHz}\), \(\sigma_h^2=1\), and \(\gamma=10\,\mathrm{dB}\) by default unless otherwise stated.

\subsection{Performance Comparisons}

\begin{table}[htb!]
\centering
\caption{Comparison of data size, encoding time, and decoding time across baseline methods (E-1: Binary Serialization, E-2: tANS \cite{duda2013asymmetric}, E-3: DietGPU \cite{dietgpu}) and our proposed approach.}
\label{tab:exp_results}
\begin{tabular}{lccc}
\toprule
\textbf{Method}         & \textbf{Data Size (KB)} & \textbf{Enc. (ms)} & \textbf{Dec. (ms)} \\
\midrule
\textbf{E-1}    & 401  & 0.139   & 0.091  \\
\textbf{E-2}        & 80   & 979.112 & 32.126 \\
\textbf{E-3}     & 156  & 1.286   & 1.117  \\
\rowcolor{lightgray}\textbf{Ours (Q=3)}  & \textbf{56}   & \textbf{0.705} & \textbf{0.479} \\
\rowcolor{lightgray}\textbf{Ours (Q=4)}  & \textbf{90}   & \textbf{0.739} & \textbf{0.517} \\
\rowcolor{lightgray}\textbf{Ours (Q=6)}  & \textbf{121}  & \textbf{0.715} & \textbf{0.494} \\
\bottomrule
\end{tabular}
\end{table}
In Table~\ref{tab:exp_results}, we compare the data size, encoding time, and decoding time of the baseline methods and our proposed method. The proposed method consistently produces smaller compressed data than the baselines. Our method, with a quantization level of Q=3, achieves a compressed size of only 56KB (7.2× lower than baseline (E-1, 401KB)), while maintaining encoding and decoding times below 1\,ms. Although tANS (E-2) \cite{duda2013asymmetric} yields a moderate compression (80KB), its encoding cost (979\,ms) rules it out for any real-time application. On the other hand, our method achieves encoding and decoding times below 1\,ms on all quantization levels (Q=3, Q=4, and Q=6), indicating its efficiency. In fact, our method achieves up to 2.8× lower data size than the optimized DietGPU (E-3) \cite{dietgpu} at the same speeds.

\begin{figure}
\centering
\includegraphics[width=\linewidth]{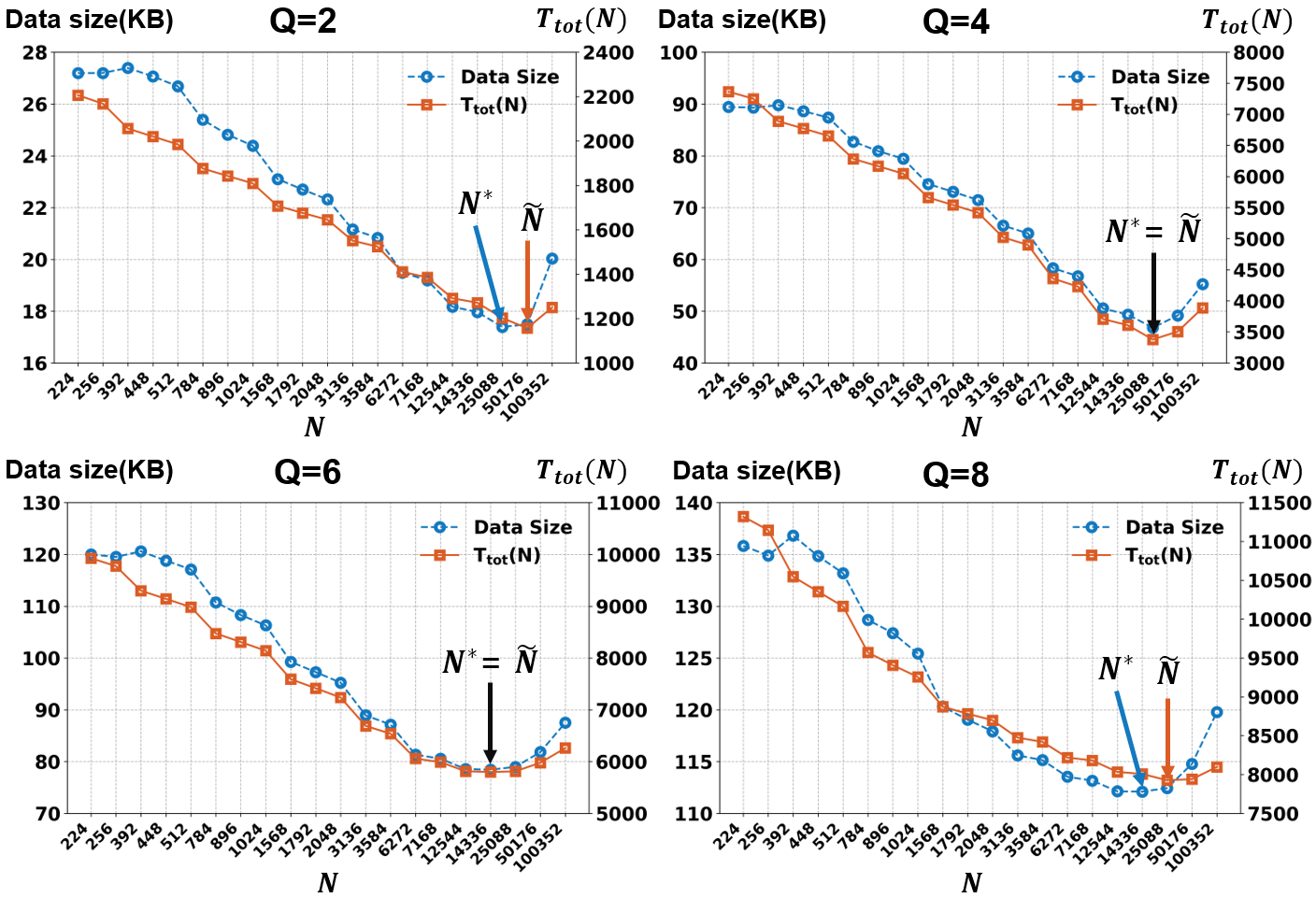} 
\caption{Results for ResNet34 with SL2 on CIFAR100 at various bit-widths ($Q=2,4,6,8$). Blue curves show compressed data sizes, while orange curves depict $T_{\mathrm{tot}}(N)$.}
\label{fig:approx_vs_actual}
\end{figure}
In Fig.~\ref{fig:approx_vs_actual}, we plot the measured $T_{\mathrm{tot}}(N)$ (solid orange lines) alongside actual data sizes (solid blue lines) and overlay our approximate cost model (dashed lines) for $Q \in \{2,4,6,8\}$. Despite minor discrepancies in absolute magnitudes, the predicted cost pattern tracks the real data closely, and the selected reshape dimension $\Tilde{N}$ reliably attains near-optimal compression efficiency.
Moreover, we also note that for Q=2, the optimal reshape dimension $N^*$ is also centrally located but at larger values than in the higher bit-width settings, indicating a consistent change of entropy features as quantization becomes more aggressive. Furthermore, our algorithm consistently identifies near-optimal reshape dimensions that lead to 2-3\% compression results from the exhaustive search global optimum.

\begin{table}[htb!]
\centering
\caption{Accuracy (\%) comparison at various quantization bit-widths.}
\label{tab:table1}
\begin{tabular}{S[table-format=1.0] S[table-format=2.2] S[table-format=2.2]}
\toprule
{$Q$} & {\textbf{ResNet34 (CIFAR100)}} & {\textbf{ResNet50 (ImageNet)}} \\
\midrule
8 & 71.27 & 74.52 \\
7 & 71.29 & 74.52 \\
6 & 71.27 & 74.51 \\
5 & 71.30 & 74.47 \\
4 & 71.29 & 74.19 \\
3 & 70.64 & 72.41 \\
2 & 66.26 & 68.14 \\
\bottomrule
\end{tabular}
\end{table}
Table~\ref{tab:table1} shows the correlation between quantization bit-width and model accuracy using ResNet34 on CIFAR100 and ResNet50 on ImageNet with a fixed split layer, SL2. For ResNet34 on CIFAR100, accuracy for the bit-width Q=4 to Q=8 remains remarkably stable (71.27-71.30\%), with degradation only at Q=3 (70.64\%). The ImageNet dataset exhibits similar robustness using the ResNet50 model,recording an accuracy of over 74\% even at Q=4.
The significant performance degradation observed for Q=2 (66.26\% and 68.14\% for CIFAR100 and ImageNet, respectively) marks a practical lower bound to quantization for real-world use cases. The results indicate that our asymmetric quantization approach reliably retains discriminative features at very low levels of precision (Q = 3,4) and can achieve aggressive compression with less impact on model performance.

\begin{table*}[ht!]
\centering
\caption{Single table combining Accuracy(\%), \(\displaystyle T_{\mathrm{comm}}(\Tilde{N})\)(ms), Data Size(Byte), and Enc/Dec times(ms) 
for Llama2 7B and 13B. In \textcolor{red}{red} we highlight how many times \(\displaystyle T_{\mathrm{comm}}(\Tilde{N}))\) is lower compared to Baseline. 
A thin horizontal rule separates Baseline from quantized rows.}
\label{tab:LLMs_comparison}
\resizebox{\textwidth}{!}{%
\begin{tabular}{@{}llcccccccccc@{}}
\toprule
& 
& \multicolumn{2}{c}{\textbf{7B: Acc. / \(T_{\mathrm{comm}}(\Tilde{N})\)}} 
& \multicolumn{2}{c}{\textbf{13B: Acc. / \(T_{\mathrm{comm}}(\Tilde{N})\)}} 
& \multicolumn{3}{c}{\textbf{7B: Size / Enc / Dec}} 
& \multicolumn{3}{c}{\textbf{13B: Size / Enc / Dec}} \\
\cmidrule(lr){3-4}\cmidrule(lr){5-6}\cmidrule(lr){7-9}\cmidrule(lr){10-12}
\textbf{Dataset} 
& \textbf{Method}
& \textbf{Acc.} 
& \(\displaystyle T_{\mathrm{comm}}(\Tilde{N})\)
& \textbf{Acc.} 
& \(\displaystyle T_{\mathrm{comm}}(\Tilde{N})\)
& \textbf{Size} 
& \textbf{Enc (mean$\pm$std)} 
& \textbf{Dec (mean$\pm$std)}
& \textbf{Size} 
& \textbf{Enc (mean$\pm$std)} 
& \textbf{Dec (mean$\pm$std)} \\
\midrule

%========================== MMLU ==========================
\multirow{5}{*}{\textbf{MMLU}} 
& Baseline 
  & 34.15 & 6.62 
  & 41.28 & 8.27 
  & 3.24M & 0.78(0.37) & 1.13(0.53) 
  & 4.05M & 0.74(0.34) & 1.10(0.51)
\\
\cmidrule(lr){2-12}
& Q=2 
  & 27.96 {\color{blue}(-6.19)} 
  & 1.56 \textcolor{red}{(4.24$\times$)} 
  & 24.57 {\color{blue}(-16.71)} 
  & 2.69 \textcolor{red}{(3.07$\times$)}
  & 0.76M & 0.74(0.02) & 0.57(0.02) 
  & 1.31M & 0.73(0.02) & 0.56(0.02)
\\
& Q=4 
  & 31.21 {\color{blue}(-2.94)} 
  & 2.49 \textcolor{red}{(2.66$\times$)} 
  & 33.30 {\color{blue}(-7.98)}
  & 3.13 \textcolor{red}{(2.64$\times$)}
  & 1.22M & 0.75(0.02) & 0.58(0.02) 
  & 1.53M & 0.73(0.02) & 0.57(0.02)
\\
& Q=6 
  & 33.50 {\color{blue}(-0.65)} 
  & 2.50 \textcolor{red}{(2.65$\times$)} 
  & 41.91 {\color{red}(+0.63)}
  & 3.14 \textcolor{red}{(2.63$\times$)}
  & 1.23M & 0.74(0.02) & 0.58(0.02) 
  & 1.54M & 0.73(0.02) & 0.57(0.02)
\\
& Q=8 
  & 34.13 {\color{blue}(-0.02)} 
  & 2.92 \textcolor{red}{(2.27$\times$)} 
  & 42.04 {\color{red}(+0.76)}
  & 3.34 \textcolor{red}{(2.48$\times$)}
  & 1.43M & 0.75(0.02) & 0.59(0.02) 
  & 1.64M & 0.73(0.02) & 0.57(0.02)
\\
\midrule

%========================== HellaSwag ==========================
\multirow{5}{*}{\textbf{HellaSwag}} 
& Baseline 
  & 73.80 & 0.60 
  & 77.25 & 0.74
  & 2.92M & 1.23(0.37) & 1.12(0.53)
  & 3.63M & 1.32(0.34) & 1.09(0.51)
\\
\cmidrule(lr){2-12}
& Q=2 
  & 72.84 {\color{blue}(-0.96)} 
  & 0.14 \textcolor{red}{(4.23$\times$)}
  & 47.40 {\color{blue}(-29.85)} 
  & 0.24 \textcolor{red}{(3.05$\times$)}
  & 0.69M & 0.73(0.02) & 0.54(0.02)
  & 1.19M & 0.72(0.02) & 0.53(0.02)
\\
& Q=4 
  & 73.90 {\color{red}(+0.10)}
  & 0.22 \textcolor{red}{(2.65$\times$)} 
  & 63.93 {\color{blue}(-13.32)} 
  & 0.28 \textcolor{red}{(2.63$\times$)}
  & 1.10M & 0.73(0.02) & 0.53(0.02)
  & 1.38M & 0.72(0.02) & 0.52(0.02)
\\
& Q=6 
  & 73.95 {\color{red}(+0.15)}
  & 0.23 \textcolor{red}{(2.63$\times$)}
  & 76.44 {\color{blue}(-0.81)}
  & 0.28 \textcolor{red}{(2.62$\times$)}
  & 1.11M & 0.73(0.02) & 0.54(0.02)
  & 1.39M & 0.72(0.02) & 0.52(0.02)
\\
& Q=8 
  & 73.81 {\color{red}(+0.01)}
  & 0.26 \textcolor{red}{(2.29$\times$)}
  & 77.29 {\color{red}(+0.04)}
  & 0.28 \textcolor{red}{(2.62$\times$)}
  & 1.27M & 0.73(0.02) & 0.54(0.02)
  & 1.42M & 0.73(0.02) & 0.53(0.02)
\\
\midrule

%========================== ARC ==========================
\multirow{5}{*}{\textbf{ARC}} 
& Baseline 
  & 53.24 & 34.79
  & 64.59 & 43.42
  & 17.06M & 1.22(0.37) & 1.25(0.53)
  & 21.29M & 1.19(0.34) & 1.21(0.51)
\\
\cmidrule(lr){2-12}
& Q=2 
  & 44.11 {\color{blue}(-9.13)} 
  & 8.16 \textcolor{red}{(4.26$\times$)}
  & 39.68 {\color{blue}(-24.91)} 
  & 14.11 \textcolor{red}{(3.08$\times$)}
  & 4.00M & 0.71(0.02) & 0.67(0.02)
  & 6.92M & 0.73(0.02) & 0.65(0.02)
\\
& Q=4 
  & 48.63 {\color{blue}(-4.61)} 
  & 13.05 \textcolor{red}{(2.67$\times$)}
  & 62.20 {\color{blue}(-2.39)}
  & 16.42 \textcolor{red}{(2.64$\times$)}
  & 6.40M & 0.70(0.02) & 0.64(0.02)
  & 8.05M & 0.74(0.02) & 0.70(0.02)
\\
& Q=6 
  & 53.84 {\color{red}(+0.60)}
  & 13.17 \textcolor{red}{(2.64$\times$)}
  & 63.14 {\color{blue}(-1.45)}
  & 16.40 \textcolor{red}{(2.65$\times$)}
  & 6.45M & 0.70(0.02) & 0.65(0.02)
  & 8.08M & 0.75(0.02) & 0.70(0.02)
\\
& Q=8 
  & 53.67 {\color{red}(+0.43)}
  & 15.39 \textcolor{red}{(2.26$\times$)}
  & 64.16 {\color{blue}(-0.43)}
  & 17.65 \textcolor{red}{(2.46$\times$)}
  & 7.55M & 0.71(0.02) & 0.66(0.02)
  & 8.65M & 0.75(0.02) & 0.71(0.02)
\\
\midrule

%========================== PIQA ==========================
\multirow{5}{*}{\textbf{PIQA}} 
& Baseline 
  & 59.58 & 0.57
  & 64.85 & 0.71
  & 0.28M & 1.17(0.37) & 1.18(0.53)
  & 0.35M & 1.23(0.34) & 1.20(0.51)
\\
\cmidrule(lr){2-12}
& Q=2 
  & 59.68 {\color{red}(+0.10)}
  & 0.14 \textcolor{red}{(4.07$\times$)}
  & 49.95 {\color{blue}(-14.90)}
  & 0.23 \textcolor{red}{(3.09$\times$)}
  & 0.07M & 0.73(0.02) & 0.53(0.02)
  & 0.11M & 0.73(0.02) & 0.53(0.02)
\\
& Q=4 
  & 58.87 {\color{blue}(-0.71)}
  & 0.22 \textcolor{red}{(2.59$\times$)}
  & 53.81 {\color{blue}(-11.04)}
  & 0.27 \textcolor{red}{(2.63$\times$)}
  & 0.11M & 0.73(0.02) & 0.51(0.02)
  & 0.13M & 0.72(0.02) & 0.52(0.02)
\\
& Q=6 
  & 59.85 {\color{red}(+0.27)}
  & 0.22 \textcolor{red}{(2.59$\times$)}
  & 64.58 {\color{blue}(-0.27)}
  & 0.27 \textcolor{red}{(2.63$\times$)}
  & 0.11M & 0.73(0.02) & 0.53(0.02)
  & 0.14M & 0.72(0.02) & 0.53(0.02)
\\
& Q=8 
  & 59.41 {\color{blue}(-0.17)}
  & 0.25 \textcolor{red}{(2.28$\times$)}
  & 65.18 {\color{red}(+0.33)}
  & 0.28 \textcolor{red}{(2.54$\times$)}
  & 0.12M & 0.74(0.02) & 0.53(0.02)
  & 0.15M & 0.72(0.02) & 0.52(0.02)
\\
\midrule

%========================== Winogrande ==========================
\multirow{5}{*}{\textbf{Winogrande}} 
& Baseline 
  & 50.43 & 4.01
  & 51.30 & 5.00
  & 1.97M & 1.01(0.37) & 1.14(0.53)
  & 2.45M & 1.02(0.34) & 1.10(0.51)
\\
\cmidrule(lr){2-12}
& Q=2 
  & 50.28 {\color{blue}(-0.15)}
  & 0.94 \textcolor{red}{(4.27$\times$)}
  & 49.57 {\color{blue}(-1.73)}
  & 1.62 \textcolor{red}{(3.09$\times$)}
  & 0.46M & 0.73(0.02) & 0.57(0.02)
  & 0.79M & 0.73(0.02) & 0.56(0.02)
\\
& Q=4 
  & 48.78 {\color{blue}(-1.65)}
  & 1.50 \textcolor{red}{(2.67$\times$)}
  & 51.46 {\color{red}(+0.16)}
  & 1.89 \textcolor{red}{(2.65$\times$)}
  & 0.74M & 0.73(0.02) & 0.58(0.02)
  & 0.93M & 0.73(0.02) & 0.56(0.02)
\\
& Q=6 
  & 50.83 {\color{red}(+0.40)}
  & 1.52 \textcolor{red}{(2.64$\times$)}
  & 50.75 {\color{blue}(-0.55)}
  & 1.90 \textcolor{red}{(2.63$\times$)}
  & 0.74M & 0.73(0.02) & 0.57(0.02)
  & 0.93M & 0.74(0.02) & 0.56(0.02)
\\
& Q=8 
  & 50.51 {\color{red}(+0.08)}
  & 1.62 \textcolor{red}{(2.48$\times$)}
  & 51.07 {\color{blue}(-0.23)}
  & 1.99 \textcolor{red}{(2.51$\times$)}
  & 0.79M & 0.73(0.02) & 0.58(0.02)
  & 0.98M & 0.73(0.02) & 0.56(0.02)
\\
\midrule

%========================== BoolQ ==========================
\multirow{5}{*}{\textbf{BoolQ}} 
& Baseline 
  & 71.13 & 22.63
  & 81.96 & 28.26
  & 11.09M & 0.67(0.37) & 0.62(0.53)
  & 13.85M & 0.67(0.34) & 0.70(0.51)
\\
\cmidrule(lr){2-12}
& Q=2 
  & 58.32 {\color{blue}(-12.81)}
  & 5.31 \textcolor{red}{(4.26$\times$)}
  & 65.93 {\color{blue}(-16.03)}
  & 9.15  \textcolor{red}{(3.09$\times$)}
  & 2.60M & 0.70(0.02) & 0.60(0.02)
  & 4.49M & 0.76(0.02) & 0.63(0.02)
\\
& Q=4 
  & 51.31 {\color{blue}(-19.82)}
  & 8.48 \textcolor{red}{(2.67$\times$)}
  & 72.78 {\color{blue}(-9.18)}
  & 10.68 \textcolor{red}{(2.65$\times$)}
  & 4.16M & 0.70(0.02) & 0.59(0.02)
  & 5.23M & 0.76(0.02) & 0.67(0.02)
\\
& Q=6 
  & 67.68 {\color{blue}(-3.45)}
  & 8.56 \textcolor{red}{(2.64$\times$)}
  & 80.86 {\color{blue}(-1.10)}
  & 10.67 \textcolor{red}{(2.65$\times$)}
  & 4.20M & 0.69(0.02) & 0.60(0.02)
  & 5.26M & 0.75(0.02) & 0.66(0.02)
\\
& Q=8 
  & 70.00 {\color{blue}(-1.13)}
  & 10.00 \textcolor{red}{(2.26$\times$)}
  & 81.65 {\color{blue}(-0.31)}
  & 10.73 \textcolor{red}{(2.63$\times$)}
  & 4.90M & 0.71(0.02) & 0.61(0.02)
  & 5.62M & 0.71(0.02) & 0.64(0.02)
\\
\midrule

%========================== OpenBookQA ==========================
\multirow{5}{*}{\textbf{OpenBookQA}}
& Baseline 
  & 57.80 & 5.03
  & 64.00 & 6.28
  & 2.47M & 0.93(0.37) & 1.10(0.53)
  & 3.07M & 0.79(0.34) & 1.27(0.51)
\\
\cmidrule(lr){2-12}
& Q=2 
  & 48.00 {\color{blue}(-9.80)} 
  & 1.18 \textcolor{red}{(4.26$\times$)} 
  & 58.40 {\color{blue}(-5.60)}
  & 2.05 \textcolor{red}{(3.06$\times$)}
  & 0.58M & 0.75(0.02) & 0.58(0.02)
  & 1.00M & 0.73(0.02) & 0.56(0.02)
\\
& Q=4 
  & 55.60 {\color{blue}(-2.20)}
  & 1.89 \textcolor{red}{(2.66$\times$)} 
  & 65.40 {\color{red}(+1.40)}
  & 2.38 \textcolor{red}{(2.64$\times$)}
  & 0.93M & 0.74(0.02) & 0.58(0.02)
  & 1.17M & 0.73(0.02) & 0.56(0.02)
\\
& Q=6 
  & 57.40 {\color{blue}(-0.40)} 
  & 1.91 \textcolor{red}{(2.63$\times$)}
  & 60.20 {\color{blue}(-3.80)}
  & 2.39 \textcolor{red}{(2.63$\times$)}
  & 0.94M & 0.74(0.02) & 0.58(0.02)
  & 1.17M & 0.74(0.02) & 0.56(0.02)
\\
& Q=8 
  & 57.40 {\color{blue}(-0.40)} 
  & 2.23 \textcolor{red}{(2.26$\times$)}
  & 63.40 {\color{blue}(-0.60)}
  & 2.45 \textcolor{red}{(2.56$\times$)}
  & 1.10M & 0.74(0.02) & 0.58(0.02)
  & 1.20M & 0.74(0.02) & 0.56(0.02)
\\
\bottomrule
\end{tabular}
}% end resizebox
\end{table*}
Table~\ref{tab:LLMs_comparison} summarizes an extensive analysis of our proposed rANS-based compression framework on LLMs, specifically comparing performance across Llama2 7B and 13B models on seven widely utilized datasets (MMLU, HellaSwag, ARC, PIQA, Winogrande, BoolQ, and OpenBookQA). Table ~\ref{tab:LLMs_comparison} presents results for accuracy, communication time, residual data size, and encoding/decoding times at different quantization levels (Q=2,4,6,8) as compared with a full-precision baseline. The proposed method achieves 2.59×-2.65× consistent transmission time reduction at Q=6 quantization while maintaining performance within ±0.65\% of the uncompressed baseline in performance for most tasks on the 7B model. Notably, task sensitivity to compression is not uniform for all tasks; PIQA and Winogrande show high robustness (±0.27\% accuracy change at Q=6), while BoolQ shows extreme vulnerability (-3.45\% at Q=6).

A remarkable model size effect appears where the Llama2 13B suffers a more significant accuracy degradation at aggressive quantization levels (Q=2, Q=4) than its Llama2 7B counterpart but fully recovers at Q=6, frequently outperforming the baseline performance (+0.63\% on MMLU). This finding implies that larger models encode more detailed representations sensitive to quantization at low levels. Still, they also exhibit enough redundancy to get away with moderate amounts of compression without a significant drop in performance.
The encoding and decoding operations are remarkably constant across all configurations (±0.02\,ms standard deviation) but counter-intuitively lower latency for compressed models (0.69-0.76\,ms) than for uncompressed baselines (0.67-1.32\,ms) because the compressed models go through less memory transfers and enjoy better cache usage. These findings establish quantization, Q=6, as the optimal operating point for bandwidth-constrained LLM deployment, providing significant communication savings without affecting model performance. However, task-specific considerations may necessitate an adjustment to Q=4 for maximal compression or Q=8 for performance-critical applications.

\begin{table}
\centering
\caption{Accuracy (\%) for different split layers (SL1--SL4).}
\label{tab:combined_table}
\begin{tabular}{
    l
    S[table-format=2.2]
    S[table-format=2.2]
    S[table-format=2.2]
    S[table-format=2.2]
}
\toprule
\multirow{2}{*}{\textbf{Split Layer}} 
& \multicolumn{2}{c}{\textbf{ResNet34 (CIFAR100)}} 
& \multicolumn{2}{c}{\textbf{ResNet50 (ImageNet)}} \\
\cmidrule(lr){2-3} \cmidrule(lr){4-5}
& {\textbf{$Q=3$}} & {\textbf{$Q=4$}} & {\textbf{$Q=3$}} & {\textbf{$Q=4$}} \\
\midrule
SL1 & 72.20 & 74.12 & 71.41 & 71.39 \\
SL2 & 72.41 & 74.19 & 70.64 & 71.29 \\
SL3 & 73.64 & 74.31 & 70.55 & 71.14 \\
SL4 & 74.66 & 74.57 & 71.44 & 71.27 \\
\bottomrule
\end{tabular}
\end{table}
Table~\ref{tab:combined_table} provides a summary of the accuracy for four split layers (SL1--SL4) for $Q=3$ and $Q=4$. The results shown in Table ~\ref{tab:combined_table} highlight the consistency of our method using distinct network split points (SL1-SL4). Indeed, for ResNet34 on CIFAR100, decent performance persists and even improves with deeper split points, from 72.20\% at SL1 to 74.66\% (at SL4 with\ Q=3). This observation implies that features deeper in the network architecture may benefit from the regularization introduced by controlled quantization. With ResNet50 results on ImageNet, the performance continues to be more stable across split points, with accuracies only differing by a maximum of 1\% across configurations. This uniformity across split points offers a lot of potential for system designers to choose where to execute edge processing or incur transmission costs depending on the specifics of data placements.

\color{black}

\begin{table}
\centering
\caption{Accuracy (\%) across diverse architectures.}
\label{tab:multi_model}
\begin{tabular}{l c c c}
\toprule
\textbf{Model} & \textbf{SL} & \textbf{Baseline} & \textbf{Ours} \\
\midrule
VGG16          & 10 & 70.200 & 70.030 (\textcolor{blue}{-0.170}) \\
MobileNetV2    & 10 & 69.858 & 69.810 (\textcolor{blue}{-0.048}) \\
SwinT          & 10 & 80.372 & 80.462 (\textcolor{red}{+0.090}) \\
DenseNet121    & 10 & 71.946 & 71.950 (\textcolor{red}{+0.004}) \\
EfficientNetB0 &  5 & 76.076 & 76.056 (\textcolor{blue}{-0.020}) \\
\bottomrule
\end{tabular}
\end{table}
To demonstrate the generality of our compression method over multiple architectures (VGG16, MobileNetV2, SwinT, DenseNet121 and EfficientNetB0), we provide ImageNet results ($Q=4$) in Table~\ref{tab:multi_model}. The results show impressive consistency across architectures, with accuracy changes $< \pm0.2\%$ of the baseline on most models. This is particularly evident with the small accuracy gains of SwinT ($+0.090\%$) and DenseNet121 ($+0.004\%$), which suggests a regularization effect of our compression pipeline for some architectures. The further consistent performance across VGG16 ($-0.170\%$), MobileNetV2 ($-0.048\%$), and EfficientNetB0 ($-0.020\%$) thus demonstrates the architecture-free characteristics of our approach. This generalizability is important for practical deployments as multiple model architectures might share the same system in a practical application.

Overall, the results of our experiments show that our rANS-based compression framework retains or even enhances overall compression efficiency while keeping model accuracy constant among various architectures, datasets, and split points. The sub-millisecond encoding and decoding times across all tests validate the real-world feasibility of our solution for latency-sensitive edge-cloud operations in high bandwidth-constrained environments.

\balance
\section{Conclusion}\label{sec:conclusion}
This paper introduced an innovative lightweight compression framework for split computing based on Range Asymmetric Numeral Systems (rANS), which jointly integrated asymmetric integer quantization and sparse CSR representation to compress intermediate features efficiently. The key contributions included an efficient probability-model-free rANS compression approach with GPU optimization for sub-millisecond latency and a theoretical framework for optimal reshape dimension selection. Through an extensive evaluation across various architectures and datasets, we demonstrated that the proposed method achieved a better compression ratio of up to 7.2× with model accuracy within 0.2\% of baselines in most cases. In addition, the proposed method provided approximate sub-millisecond encoding/decoding times, allowing real-time applications with very stringent latency requirements. Moreover, the method performed well in vision and language spaces, validating its modality-agnosticity.

Perhaps most importantly, this study expanded split computing from traditional computer vision applications to language models, yielding a 2.59×-2.65× reduction in communication times for Llama2 models while maintaining their semantic capabilities. This advance allowed for collaborative intelligence scenarios leveraging LLMs on low-resource devices, which was not feasible. The consistent performance across various architectures, split points, and modalities showed the potential of the proposed method as a versatile building block for future distributed intelligence systems.

Future work will explore dynamic bit-width adaptation according to network conditions, extended support for running multimodal models, and integration with complementary techniques such as model pruning and knowledge distillation to maximize the benefits of edge-cloud collaboration.

\bibliographystyle{ieeetr}
\bibliography{references}

\end{document}